\documentclass[aps,prd,onecolumn,groupedaddress,showpacs,nofootinbib,amssymb]{revtex4}
\usepackage[dvips]{graphicx}
\usepackage{amssymb}
\usepackage{amsmath}
\usepackage{graphicx,,color}
\usepackage{amsfonts}
\usepackage{bm}
\usepackage{cancel}
\usepackage{comment}

\newcommand\be{\begin{equation}}
\newcommand\ee{\end{equation}}

\allowdisplaybreaks[4]

\begin{document}

\tolerance=5000

\title{A Nearly Massless Graviton in Einstein-Gauss-Bonnet Inflation with Linear Coupling Implies Constant-roll for the Scalar Field}
\author{V.K.~Oikonomou,$^{1,2,3}$\,\thanks{v.k.oikonomou1979@gmail.com}F.P.
Fronimos,$^{1}$\,\thanks{fotisfronimos@gmail.com}}
\affiliation{$^{1)}$ Department of Physics, Aristotle University
of Thessaloniki, Thessaloniki 54124,
Greece\\
$^{2)}$ Laboratory for Theoretical Cosmology, Tomsk State
University of Control Systems and Radioelectronics, 634050 Tomsk,
Russia (TUSUR)\\
$^{3)}$ Tomsk State Pedagogical University, 634061 Tomsk,
Russia\\}

\tolerance=5000

\begin{abstract}
The striking GW170817 event indicated that the graviton is nearly
massless, since the gamma rays emitted from the two neutron stars
merging arrived almost simultaneously with the gravitational
waves. Thus, the graviton must also be massless during the
inflationary and post-inflationary era, since there is no obvious
reason to believe otherwise. In this letter we shall investigate
the theoretical implications of the constraint that the graviton
is massless to an Einstein-Gauss-Bonnet theory with linear
coupling of the scalar field to the four dimensional Gauss-Bonnet
invariant. As we show, the constraint of having gravitational wave
speed of the primordial gravitational waves equal to one, severely
restricts the dynamics of the scalar field, imposing a direct
constant-roll evolution on it. Also, as we show, the spectral
index of the primordial scalar perturbations for the
GW170817-compatible Einstein-Gauss-Bonnet theory with linear
coupling is different in comparison to the same theory with
non-linear coupling. Thus the phenomenology of the model is
expected to be different, and we briefly discuss this issue too.
In addition, the constant-roll condition is always related to
non-Gaussianities, thus it is interesting that the imposition of a
massless graviton in an Einstein-Gauss-Bonnet theory with linear
coupling may lead to non-Gaussianities, so we briefly discuss this
issue too.
\end{abstract}


\maketitle

\section{Introduction}

The precision cosmology era that we live has the advantage that
exist numerous observational data coming from low redshift
sources, the Cosmic Microwave Background and also from
astrophysical sources. These data shape literally our perception
of the currently accelerating Universe, and alter in some cases
the way we thought that the Universe works. One of these events in
the 2017 observation of gravitational waves coming from the
merging of two neutron stars, known now as the GW170817 event
\cite{GBM:2017lvd}. This event was astonishing because it provided
useful information for theoretical cosmologists, since the gamma
rays arrived almost simultaneously with the gravitational waves.
This result indicated that the gravitational wave speed $c_T$ is
nearly equal to that of light, that is, $c_T^2\simeq 1$, in
natural units. In effect, the graviton, which is the propagator
and mediator of gravity, is nearly massless.

From a theoretical point of view there is no reason to believe
that during the early Universe, the graviton should have different
mass from what it has today, therefore the implication of the
GW170817 is that the graviton should have nearly zero mass during
the inflationary and post-inflationary era. This requirement has a
dramatic effect for quite a number of modified gravities
describing the early Universe which predict a non-zero mass for
the graviton, since these are ruled out by the GW170817 event. For
a comprehensive account on this topic see for example
\cite{Ezquiaga:2017ekz}.

However, most of the fundamental modified gravities like $f(R)$
gravity or Gauss-Bonnet gravity
\cite{Nojiri:2017ncd,Nojiri:2010wj,Nojiri:2006ri,Capozziello:2011et,Capozziello:2010zz,delaCruzDombriz:2012xy,Olmo:2011uz},
still predict a massless graviton even at the primordial era, thus
are not ruled out. Nevertheless, an interesting class of modified
gravities, namely Einstein-Gauss-Bonnet gravities
\cite{Hwang:2005hb,Nojiri:2006je,Cognola:2006sp,Nojiri:2005vv,Nojiri:2005jg,Satoh:2007gn,Bamba:2014zoa,Yi:2018gse,Guo:2009uk,Guo:2010jr,Jiang:2013gza,Kanti:2015pda,vandeBruck:2017voa,Kanti:1998jd,Kawai:1999pw,Nozari:2017rta,Odintsov:2018zhw,Kawai:1998ab,Yi:2018dhl,vandeBruck:2016xvt,Kleihaus:2019rbg,Bakopoulos:2019tvc,Maeda:2011zn,Bakopoulos:2020dfg,Ai:2020peo,Easther:1996yd,Antoniadis:1993jc,Antoniadis:1990uu,Kanti:1995vq,Kanti:1997br,Odintsov:2019clh,Odintsov:2020sqy,Odintsov:2020zkl},
became quite problematic, since the predicted primordial
gravitational wave speed is non-zero, thus the tensor
perturbations of the primordial Universe produce results
incompatible with the GW170817 event. In view of these problem, in
some previous works
\cite{Odintsov:2019clh,Odintsov:2020sqy,Odintsov:2020zkl} we
investigated under which conditions it is possible to obtain a
massless graviton in the context of Einstein-Gauss-Bonnet
theories. As we demonstrated, the requirement that $c_T^2\simeq 1$
is obtained only when the coupling function $\xi (\phi)$ of the
scalar field to the four dimensional Gauss-Bonnet invariant
satisfies the differential equation $\ddot\xi-H\dot\xi=0$. In
Refs. \cite{Odintsov:2020sqy,Odintsov:2020zkl} we showed that the
quoted differential equation can severely constrain the functional
forms of the coupling function $\xi (\phi)$ and of the scalar
potential $V(\phi)$. As we showed, the viability of the
GW170817-compatible Einstein-Gauss-Bonnet models can be achieved
if the slow-roll conditions hold true. Thus we can simultaneously
obtain a viable inflationary era from an Einstein-Gauss-Bonnet
theory that predicts massless gravitons.

However, in Refs. \cite{Odintsov:2020sqy,Odintsov:2020zkl} we did
not take into account at all the case that the coupling function
$\xi (\phi)$ is a linear function of the scalar field. The purpose
of this letter is to address exactly this case, and to investigate
the effects of the linear scalar coupling function on the
inflationary phenomenology of the GW170817-compatible
Einstein-Gauss-Bonnet theories. As we show, it has dramatic
effects, since it imposes a fixed constant-roll evolution on the
scalar field thus utterly affecting the scalar field evolution. We
focus on the behavior of the spectral index of the scalar
perturbations, and we explicitly demonstrate how the linear
coupling function changes it. As we show, the phenomenology is
changed in comparison to the cases that the coupling function $\xi
(\phi)$ is non-linear, which we studied in Refs.
\cite{Odintsov:2020sqy,Odintsov:2020zkl}. This result affects many
theoretical frameworks which use linear functions of the scalar
field coupled to the four dimensional Gauss-Bonnet invariant, see
for example \cite{Kanti:1995vq,Carson:2020ter}, and also see Ref.
\cite{Capozziello:2008gu} for a non-local Gauss-Bonnet gravity
theory which is equivalent to a particular potential-less
Einstein-Gauss-Bonnet gravity with linear coupling of the scalar
field to the Gauss-Bonnet invariant (for a similar study in the
context of $f(R)$ gravity see for example \cite{Nojiri:2019dio}).

\section{Linear Coupling GW170817-Einstein Gauss-Bonnet Gravity:  How Constant-roll Evolution for the Scalar Field is Imposed}

Let us demonstrate how the inflationary dynamics of the scalar
field is affected if the GW170817 compatible Einstein-Gauss-Bonnet
gravity has a linear coupling function $\xi(\phi)$. Let us start
with the gravitational action of the Einstein-Gauss-Bonnet
gravity, which is,
\begin{equation}
\centering \label{action}
S=\int{d^4x\sqrt{-g}\left(\frac{R}{2\kappa^2}-\frac{\omega}{2}g^{\mu\nu}\partial_\mu\phi\partial_\nu\phi-V(\phi)-\xi(\phi)\mathcal{G}\right)}\,
,
\end{equation}
with $R$ denoting the Ricci scalar, $\kappa=\frac{1}{M_P}$ being
the gravitational constant with $M_P$ being the reduced Planck
mass, and $V(\phi)$ is the scalar potential. Also $\xi(\phi)$ is
the coupling function to the Gauss-Bonnet invariant, which is
equal to
$\mathcal{G}=R_{\mu\nu\sigma\rho}R^{\mu\nu\sigma\rho}-4R_{\mu\nu}R^{\mu\nu}+R^2$,
where $R_{\mu\nu}$ and $R_{\mu\nu\sigma\rho}$ are the Ricci and
Riemann curvature tensor respectively. In addition, we shall
assume a flat Friedman-Robertson-Walker (FRW) background with line
element,
\begin{equation}
\centering \label{metric}
ds^2=-dt^2+a^2(t)\sum_{i=1}^{3}{(dx^i)^2}\, ,
\end{equation}
where as usual $a(t)$ is the scale factor of the Universe. For the
flat FRW metric, the Ricci scalar and the Gauss-Bonnet invariant
take quite simple forms and these are, $R=6(2H^2+\dot H)$ and
$\mathcal{G}=24H^2(H^2+\dot H)$ where $H=\frac{\dot a}{a}$ denotes
the Hubble rate and as usual, the ``dot'' implies differentiation
with respect to cosmic time $t$.

The speed of the cosmological tensor perturbations for an
Einstein-Gauss-Bonnet theory is not equal to that of light, but it
is equal to \cite{Hwang:2005hb}
\begin{equation} \centering
\label{cT} c_T^2=1-\frac{Q_f}{2Q_t}\, ,
\end{equation}
where $Q_f=16(\ddot\xi-H\dot\xi)$,
$Q_t=\frac{1}{\kappa^2}-8\dot\xi H$. Thus in order to comply with
the GW170817 results, by imposing the condition $c_T^2=1$, a
constraint which would make the gravitational wave speed equal to
unity in natural units, and equal to that of the speed of light,
this leads to the constraint $Q_f=0$, or equivalently, to the
differential equation,
\begin{equation}\label{diffeeqn}
\ddot\xi=H\dot\xi\, ,
\end{equation}
which must be satisfied by the coupling function. Thus by
expressing the cosmic time derivatives with respect to the scalar
field, by using $\frac{d}{dt}=\dot\phi\frac{d}{d\phi}$, we have
$\dot\xi=\xi'\dot\phi$ and thus we can rewrite the differential
equation (\ref{diffeeqn}) as follows,
\begin{equation}
\label{constraint1} \centering
\xi''\dot\phi^2+\xi'\ddot\phi=H\xi'\dot\phi\, .
\end{equation}
Let us now get to the core of this work, which is based on the
choice of a linear coupling function, namely,
\begin{equation}\label{linearcoupling}
\xi (\phi)=c_1\kappa\phi\, ,
\end{equation}
where $c_1$ is a dimensionless constant. Thus, for the linear
coupling choice, the term containing $\xi''$ is eliminated from
Eq. (\ref{constraint1}), thus by substituting $\xi(\phi)$ from Eq.
(\ref{linearcoupling}) in Eq. (\ref{constraint1}) we get,
\begin{equation}
\centering \label{dotphi} \ddot\phi=H\dot\phi\, .
\end{equation}
The above condition describes an exact constant-roll evolution for
the scalar field, and in fact with a very specific rate. Let us
see how this affects the inflationary phenomenology of the
GW170817 Einstein-Gauss-Bonnet theory with linear coupling. To
this end, let us recall the definition of the slow-roll indices
for the Einstein-Gauss-Bonnet theory \cite{Hwang:2005hb},
\begin{align}
\centering \label{indices} \epsilon_1&=-\frac{\dot
H}{H^2}&\epsilon_2&=\frac{\ddot\phi}{H\dot\phi}&\epsilon_4&=\frac{\dot
E}{2HE}&\epsilon_5&=\frac{Q_a}{2HQ_t}&\epsilon_6&=\frac{\dot
Q_t}{2HQ_t}\, ,
\end{align}
with $Q_a=-8c_1\kappa\dot\phi H^2$ and
$E=\frac{\omega}{\kappa^2}+\frac{3Q_a^2}{2Q_t\kappa^2\dot\phi^2}$.
By using Eq. (\ref{dotphi}) and substituting it in the slow-roll
$\epsilon_2$ in Eq. (\ref{indices}) we get $\epsilon_2=1$. This
condition can affect significantly the inflationary phenomenology
of Einstein-Gauss-Bonnet theory as we now show. We shall mainly
focus on the spectral index of the primordial scalar curvature
perturbations, since this observable will be mainly affected by
the constant-roll condition (\ref{dotphi}). Let us recall how the
spectral index can be extracted from the scalar curvature
perturbations, and following \cite{Hwang:2005hb}, the calculation
of the power spectrum results to the following expression for the
parameter $z$,
\begin{align}\label{firstbigequation}
&
\frac{z''}{z}=a_c^2H_c^2(1+\epsilon_1+\epsilon_2+\epsilon_4)(2+\epsilon_2+\epsilon_4)+a^2H(\dot{\epsilon}_1+\dot{\epsilon}_2+\dot{\epsilon}_4)\\
\notag &
-2a_c^2H_c\left(\frac{3}{2}+\epsilon_1+\epsilon_2+\epsilon_4\right)\frac{\dot{\epsilon}_5}{1+\epsilon_5}-\frac{a^2\ddot{\epsilon}_5}{1+\epsilon_5}+\frac{2a^2\dot{\epsilon}_5^2}{(1+\epsilon_5)^2}\,
,
\end{align}
where $z$ is defined as follows,
\begin{equation}\label{perturbationeqnsparameters}
z=\frac{a}{(1+\epsilon_3)H}\sqrt{E}\, ,
\end{equation}
with $a$ being the scale factor, and the parameter $E$ was defined
previously below Eq. (\ref{indices}). Also $a_c$ and $H_c$ denote
the scale factor and the Hubble rate exactly at the first horizon
crossing time instance, respectively, and furthermore all the
slow-roll indices have to be evaluated at the first horizon
crossing time instance. As it was shown in Ref.
\cite{Oikonomou:2020krq}, by making use of the Karamata's theorem
we can write,
\begin{equation}\label{etarelation}
\eta=-\frac{1}{a_cH_c}\frac{1}{1-\epsilon_1}\, .
\end{equation}
where $\eta$ is the conformal time, and recall that $a_c$ and
$H_c$ denote the scale factor and the Hubble rate exactly at the
first horizon crossing time instance. Thus, by making use of Eq.
(\ref{etarelation}) in conjunction with the fact that
$\epsilon_2=1$, and by assuming that the rest of the slow-roll
indices and their derivatives satisfy the slow-roll conditions
$\epsilon_i\ll 1$ and $\dot{\epsilon_i}\ll 1$ with $i=1,4,5,6$, we
obtain the following expression from Eq. (\ref{firstbigequation}),
\begin{equation}\label{derivatives1}
\frac{z''}{z}=\frac{n_s}{\eta^2}=\frac{1}{\eta^2}\frac{1}{(1-\epsilon_1)^2}\left(2+\epsilon_1+\epsilon_4
\right)\left(3+\epsilon_4 \right)\, ,
\end{equation}
with the parameter $n_s$ being defined in the following way,
\begin{equation}\label{nsparameter}
n_s=\frac{1}{(1-\epsilon_1)^2}\frac{1}{(1-\epsilon_1)^2}\left(2+\epsilon_1+\epsilon_4
\right)\left(3+\epsilon_4 \right)\, ,
\end{equation}
and $n_s$ should not be confused with the spectral index $n_S$. By
using the fact that we assumed $\epsilon_i\ll 1$, the expression
for $n_s$ given in Eq. (\ref{nsparameter}) can be further
simplified by keeping only linear order terms in terms of the
slow-roll indices, as follows,
\begin{equation}\label{finalns}
n_s=\frac{1}{(1-\epsilon_1)^2}\left(6+3\epsilon_1-5\epsilon_3+5\epsilon_4
\right)\, .
\end{equation}
The definition of the spectral index $n_{\mathcal{S}}$ in terms of
the parameter $n_s$ is \cite{Nojiri:2017ncd,Hwang:2005hb},
\begin{equation}\label{spectralindexprofinal}
n_{\mathcal{S}}=4-\sqrt{4n_s+1}\, .
\end{equation}
Hence, by substituting Eq. (\ref{finalns}) in Eq.
(\ref{spectralindexprofinal}) we obtain the following expression
for the spectral index of the scalar primordial perturbations,
\begin{equation}\label{nssemifinal}
n_S=4-\frac{\sqrt{25+10(\epsilon_1+2\epsilon_4)}}{1-\epsilon_1}\,
,
\end{equation}
so by setting $x=\epsilon_1+2\epsilon_4$ and since $\epsilon_1\ll
1$ and $\epsilon_4\ll 1$, this means that $x\ll 1$, therefore by
expanding (\ref{nssemifinal}) for $x\ll 1$ we get,
\begin{equation}\label{semifinal}
n_S\simeq \frac{5 \epsilon_1+2 \epsilon_4+1}{\epsilon_1-1}\, .
\end{equation}
so by further expanding the above for $\epsilon_1\ll 1$ and by
keeping linear terms containing the slow-roll indices in the final
expressions, we obtain at leading order,
\begin{equation}\label{nsfinal}
n_S\simeq -1-6 \epsilon_1-2 \epsilon_4\, .
\end{equation}
Therefore, it is apparent from the above that the resulting
phenomenology for an GW170817-compatible Einstein-Gauss-Bonnet
theory with a linear coupling is quite different from the
resulting phenomenology of the same theories with non-linear
coupling, in which case the spectral index is equal to
\cite{Hwang:2005hb,Odintsov:2020sqy,Odintsov:2020zkl},
\begin{equation}\label{nsordinarytheories}
n_S=1-4 \epsilon_1-2 \epsilon_2-\epsilon_4\, .
\end{equation}
This result may cast some doubt on the phenomenology of the
Einstein-Gauss-Bonnet models with linear coupling function, since
in order to obtain a viable phenomenology it is required that
$\epsilon_1<0$ and $\epsilon_1\sim \mathcal{O}(10^{-1})$. However,
the Planck data \cite{Akrami:2018odb} constrain $\epsilon_1$ to be
of the order $\epsilon_1\sim \mathcal{O}(10^{-3})$. The Planck
results however are based on single scalar field cosmology, so the
viability of the GW170817-compatible Einstein-Gauss-Bonnet theory
with linear coupling must be checked explicitly, a task we aim to
address comprehensively in the near future. The resulting
phenomenology may be possibly affected by the presence of the
scalar field potential and our study covers many
Einstein-Gauss-Bonnet theories appearing in the literature with
linear coupling, see for example
\cite{Kanti:1995vq,Carson:2020ter}.

Another important class of theories covered by our study, is that
of non-local Gauss-Bonnet gravities \cite{Capozziello:2008gu}, in
which case the gravitational action is,
\begin{equation}
\centering
\label{action2}
S=\int{d^4x\sqrt{-g}\left(\frac{R}{2\kappa^2}-\frac{\kappa^2}{2\omega}\mathcal{G}\square^{-1}\mathcal{G}\right)}\, ,
\end{equation}
where again $R$ is the Ricci scalar, $\kappa=\frac{1}{M_P}$ and
$\mathcal{G}=R^2-4R_{\mu\nu}R^{\mu\nu}+R_{\mu\nu\sigma\rho}R^{\mu\nu\sigma\rho}$.
This gravitational action corresponds to a non-local Gauss-Bonnet
gravity, which as was shown in Ref. \cite{Capozziello:2008gu},
corresponds to the following Einstein-Gauss-Bonnet theory with
linear coupling,
\begin{equation}
\centering
\label{action2B}
S=\int{d^4x\sqrt{-g}\left(\frac{R}{2\kappa^2}-\frac{\omega}{2\kappa^2}g^{\mu\nu}\partial_{\mu}\phi\partial_\nu\phi+\phi\mathcal{G}\right)}\, ,
\end{equation}
which can be obtained from the non-local action by using the
transformation
$\phi=-\frac{\kappa^2}{\omega}\square^{-1}\mathcal{G}$. In this
case, the scalar potential is absent and the scalar field itself
is dimensionless, in contrast to the previous gravitational action
studied in this paper, namely the action of Eq. (\ref{action}).
The same cosmological background shall be assumed, meaning that
Eq. (\ref{metric}) still applies and also, the scalar field now
is, due to its definition, only time dependent, exactly as assumed
before. The important characteristic is that the Gauss-Bonnet
coupling still remains a linear function of the scalar field. As a
result, compatibility with the recent GW170817 can be once again
restore by assuming that $\ddot\phi=H\dot\phi$, hence the
constant-roll assumption is still applicable. The phenomenology of
this model, which has no scalar potential, and unusual physical
dimensions for the scalar field, shall also be developed in a
forthcoming work.

\section{Conclusions}

In this letter we investigated the possibility of obtaining a
massless graviton during the inflationary era, from an
Einstein-Gauss-Bonnet theory with a linear coupling of the scalar
field to the Gauss-Bonnet invariant. As we demonstrated, the
implication of having gravitational wave speed $c_T^2=1$ during
the inflationary era, restricts the dynamical evolution of the
scalar field and it compels it to evolve dynamically in a
constant-roll way. Accordingly we calculated the spectral index of
scalar primordial perturbations, and as we showed, the resulting
expression of it in terms of the slow-roll indices is quite
different in comparison to the GW170817-compatible
Einstein-Gauss-Bonnet theories with general coupling of the scalar
field to the Gauss-Bonnet invariant.

In principle, the resulting phenomenology of the
GW170817-compatible Einstein-Gauss-Bonnet theories with linear
coupling is seriously affected, and this should be explicitly
checked. The most serious problem is that the first slow-roll
index, which quantifies the fact that inflation occurs since it
requires $\dot{H}\ll H^2$, must be of the order $\epsilon_1\sim
\mathcal{O}(10^{-1})$ in order to have viability with the Planck
data, however, the Planck data indicate that it should be of the
order $\epsilon_1\sim \mathcal{O}(10^{-3})$. The Planck data
however are based on single scalar field inflation considerations,
so in a forthcoming work we aim to address the phenomenology of
the GW170817-compatible Einstein-Gauss-Bonnet theories with linear
coupling in detail.

Another issue that is worth investigating is the non-Gaussianities
predicted by the GW170817-compatible Einstein-Gauss-Bonnet theory
with linear coupling. The constant-roll dynamics for the scalar
field which is inherent to this theory, is known to produce
primordial non-Gaussianities \cite{Odintsov:2019ahz}, see also
Ref. \cite{Martin:2012pe} for the canonical scalar field cases.
Thus it is interesting that the imposition of having a massless
graviton for the Einstein-Gauss-Bonnet theory with linear
coupling, leads to a theory that may have non-Gaussianities. Thus
it is worth to pursuit this research issue further through the
prism of primordial non-Gaussianities, which we aim to perform in
a near future study.

\end{document}